\documentclass[
aip,
amsmath,amssymb,
preprint,
]{revtex4-2}

\usepackage{natbib}
\usepackage{graphicx}
\usepackage{hyperref}
\usepackage{cleveref}
\usepackage{fancyhdr}
\usepackage[absolute, overlay]{textpos}
\usepackage{xcolor}
\usepackage{float}

\newcommand{\startcolor}{\color{black}}
\newcommand{\bluecolor}{\color{black}}
\newcommand{\stopcolor}{\color{black}}

\begin{document}

\begin{textblock*}{20cm}(3.5cm,26cm) 
   Distribution Statement A. Approved for public release: Distribution is unlimited
\end{textblock*}

\begin{textblock*}{20cm}(3cm,25cm) 
   $^*$Author to whom correspondence should be addressed: shelby.s.fields.civ@us.navy.mil
\end{textblock*}

\title{Facile Optimization of Combinatorial Sputtering Processes with Arbitrary Numbers of Components for Targeted Compositions}

\author{Shelby Sutton Fields$^*$}
\affiliation{Materials Science and Technology Division, U.S. Naval Research Laboratory, Washington D.C., 20375}

\author{Christopher David White}
\affiliation{Materials Science and Technology Division, U.S. Naval Research Laboratory, Washington D.C., 20375}

\author{Keith E. Knipling}
\affiliation{Materials Science and Technology Division, U.S. Naval Research Laboratory, Washington D.C., 20375}

\author{Steven P. Bennett}
\affiliation{Materials Science and Technology Division, U.S. Naval Research Laboratory, Washington D.C., 20375}

\date{\today}

\begin{abstract}

Combinatorial sputtering is a physical vapor deposition method that enables the high-throughput synthesis of compositionally varied thin films. Using this technique, the effects of stoichiometry on specific properties of alloy thin films with analog composition gradients can be mapped using high-throughput characterization. To obtain specific stoichiometries, such as those desired for an equiatomic, intermetallic, or doped compounds, the sputter power of each target must be simultaneously tuned to optimize the deposition rate of each component. This optimization problem increases in complexity with the number of components, which commonly leads to iterative guess-and-check processing and can limit the intrinsic high-throughput advantages of this synthesis method. To circumvent this challenge, this work introduces a composition optimization procedure that enables the facile synthesis of sputtered combinatorial films with targeted compositions. This procedure leverages the expeditious mapping of composition using wavelength dispersive x-ray fluorescence and is capable of optimizing processing for an arbitrary number of components. As a demonstration, this method is leveraged to sputter a combinatorial Cr$_{v}$Fe$_{w}$Mo$_{x}$Nb$_{y}$Ta$_{z}$ film with an equiatomic composition near the wafer center.

\end{abstract}

\maketitle

\section{Introduction}

    Combinatorial deposition is a family of synthesis methods for coatings and thin films (referred to as ${\it{libraries}}$) that achieve compositional spreads through asymmetric processing conditions.\cite{mcginn_thin-film_2019} These methods employ different parent deposition techniques, such as magnetron sputtering,\cite{siol_combinatorial_2016} electron-beam evaporation,\cite{garcia_combinatorial_2007} pulsed laser deposition,\cite{snyder_material_2015} chemical solution deposition,\cite{zakay_combinatorial_2023} and molecular beam epitaxy\cite{logvenov_combinatorial_2007} in combination with non-equilibrium synthesis conditions, such as annealing temperature gradients\cite{sasaki_identifying_2020} or tilted material sources.\cite{fields_high-throughput_2025,liu_combinatorial_2022} Materials that have been investigated and optimized using this family of techniques include multicomponent catalysts,\cite{choi_combinatorial_2015} superconductors,\cite{jin_combinatorial_2013,logvenov_combinatorial_2007} thermoelectrics,\cite{snyder_material_2015} metallic glasses,\cite{li_high-temperature_2019} and magnetic alloys,\cite{iwasaki_machine_2021,fackler_combinatorial_2017} among others.\cite{gebhardt_combinatorial_2012,shi_high-throughput_2020,siol_combinatorial_2016,green_applications_2013,kauffmann_combinatorial_2017}

    Combinatorial processing enables high throughput material exploration and optimization due to the large spread of compositions made available for analysis in a single synthesis step. In the case of combinatorial film deposition, access to all analog compositions is afforded by property mapping, which presents a throughput advantage over methods that rely on the discrete synthesis of individual samples, which produce digital compositions.\cite{mcginn_thin-film_2019} Accordingly, characterization methods to probe properties such as optical reflectivity,\cite{schenck_high_2004} magnetic coercive field,\cite{fields_high-throughput_2025} and thermoelectric figure of merit,\cite{otani_high-throughput_2007} among others,\cite{al_hasan_combinatorial_2020,nagy_mapping_2024} have been adapted to map across large substrates and take full advantage of the available composition space. For example, the corrosion resistance properties of AlCoCrFeNi have recently been explored using combinatorial sputter deposition in combination with a droplet-cell corrosion system adapted for wafer mapping.\cite{sur_high_2023} Similarly, a modified electorchemical cell has recently been employed to investigate the catalytic properties of CoNiTi and AuFeNi alloys for hydrogen evolution reactions.\cite{liu_combinatorial_2022}
    
    For the combinatorial investigation of specific applications or materials systems, it is commonly necessary to tune deposition processes to obtain specific composition windows.\cite{marshal_combinatorial_2017} In this regard, magnetron sputter deposition posesses the advantage of a straightforward relationship between growth rate and applied target power.\cite{thelen_python-based_2025} Moreover, most commercially available dedicated sputter systems are naturally capable of simultaneous deposition from several targets, making them easily modified for combinatorial investigations.\cite{fields_high-throughput_2025} While the synthesis of binary combinatorial films with desirable compositions is straightforward, the increased complexity with additional components can make process calibration nontrivial. For example, tuning deposition conditions to achieve specific equiatomic compositions requires either process iteration\cite{nagy_mapping_2024} or an $\it{a}$ $\it{priori}$ knowledge of the relative deposition/growth rates of each component including imposed asymmetric conditions. Similar difficulties are encountered when optimizing combinatorial processes to achieve dilute compositions of one or several specific species.\cite{fields_high-throughput_2025,nadaud_enhancement_2024} Such challenges have precipitated the recent development of combinatorial systems and processes that enable the efficient\cite{snyder_material_2015,thelen_python-based_2025} or real-time\cite{suram_combinatorial_2015} optimization of film composition, enhancing throughput and enabling easy access to relevant film stoichiometry spaces.

    Within this work, a calibration procedure is introduced that enables the facile synthesis of sputtered combinatorial thin films with targeted compositions and an arbitrary number of components. This method leverages wavelength-dispersive x-ray flouresence (WDXRF) mapping, is adaptable to different wafer sizes and substrate materials without the need for re-calibration, is compatible with different composition measurement techniques, and can suggest multiple conditions to achieve the same compositions. As a demonstration, this procedure is followed to prepare a combinatorial CrFeMoNbTa alloy thin film on a 6-inch Si wafer that displays an equiatomic composition near wafer center, which is verified using energy-dispersive x-ray spectroscopy (EDS) measurements. This process enhances the throughput and accuracy of sputtered combinatorial synthesis, and is enabling for the rapid investigation of the properties of thin film alloys with complex compositions.

\section{Experimental Methods}

\subsection{Combinatorial Deposition}

    Combinatorial direct-current sputter deposition of single (Mo, Cr) and multicomponent (Cr$_{v}$Fe$_{w}$Mo$_{x}$Nb$_{y}$Ta$_{z}$, $v$, $w$, $x$, $y$, and $z$ $\leq$ 1) alloy thin films were conducted within an AJA ATC ORION deposition system equipped with several AJA A315-LP 1.5-inch sputter guns set in a custom flange that angled each toward substrate edges. The base pressure for this system was 5 $\times 10^{-8}$ Torr, and all depositions were conducted in a 4 mTorr Ar working gas \startcolor ($>$ 99.999 \% purity) \stopcolor obtained by constraining a 30 sccm mass flow-controlled feed using an automated VAT gate valve. Each individual target material was deposited using a corresponding gun and power supply, as detailed in supplemental Table S1, \startcolor following five minutes of presputter surface conditioning at deposition power\stopcolor. 6-inch diameter Si wafers were employed as substrates for each deposition. Each wafer was loaded into the \bluecolor vented \stopcolor deposition chamber without surface treatment \bluecolor before pulling the vacuum down to base pressure\stopcolor. A custom stage was employed that held the wafer at a height of 9 cm above the 1.5-inch diameter targets, which were tilted 9.5$^{\circ}$ away from vertical.\cite{fields_high-throughput_2025} The chamber posesses five total guns that are equally seperated by a radial angle of 72$^{\circ}$. All depositions were conducted by imposing direct current (DC) powers of between 2.6 and 6.6 W cm$^{-2}$ for 60 minutes.  \startcolor Targets for each material were obtained from ACI alloys and specified for $>$ 99.5 at. \% purity.\stopcolor

\subsection{Combinatorial Characterization}

    Each deposited wafer was subjected to WDXRF composition mapping using a Rigaku ZSX Primus 400 system equipped with a LiF (200)-oriented crystal coupled to a proportional count detector. Composition map grids were collected within the central $\pm$ 6.75 cm of each wafer (to avoid edge effects that would affect composition normalization) with a spot size of 1 cm that was incremented at distances of 0.75 cm on center allowing for 0.25 cm overlap between neighboring measurements. Each species map required 2 hours of measurement time for each 6 inch wafer, and a thin film model was assumed for acquisition and analysis \startcolor within the ZSX Guidance (version 8.132) software, which reports compositions in units of areal mass density\stopcolor. All WDXRF composition profile fitting \startcolor and analysis \stopcolor was completed using the lmfit (version 1.1.3) module in python (version 3.11.11).

\subsection{Cross Technique Composition Verification}

    Select positions on a combinatorial film deposited using optimized powers were characterized using EDS within a Thermoscientific Quattro S scanning electron microscope (SEM). This sytem was equipped with an AMETEK Octane Super 60 mm$^2$ detector, and EDS spectra were collected with a 15 kV accelerating voltage and analyzed with AMETEK APEX commercial analysis software without specific standards. Each measurement was integrated over a 100 $\mu$m field-width area for 1 minute.

\section{Results}
\subsection{Single-Component Deposition Characteristics}

\begin{figure*}
    
        \includegraphics[width=1\textwidth]{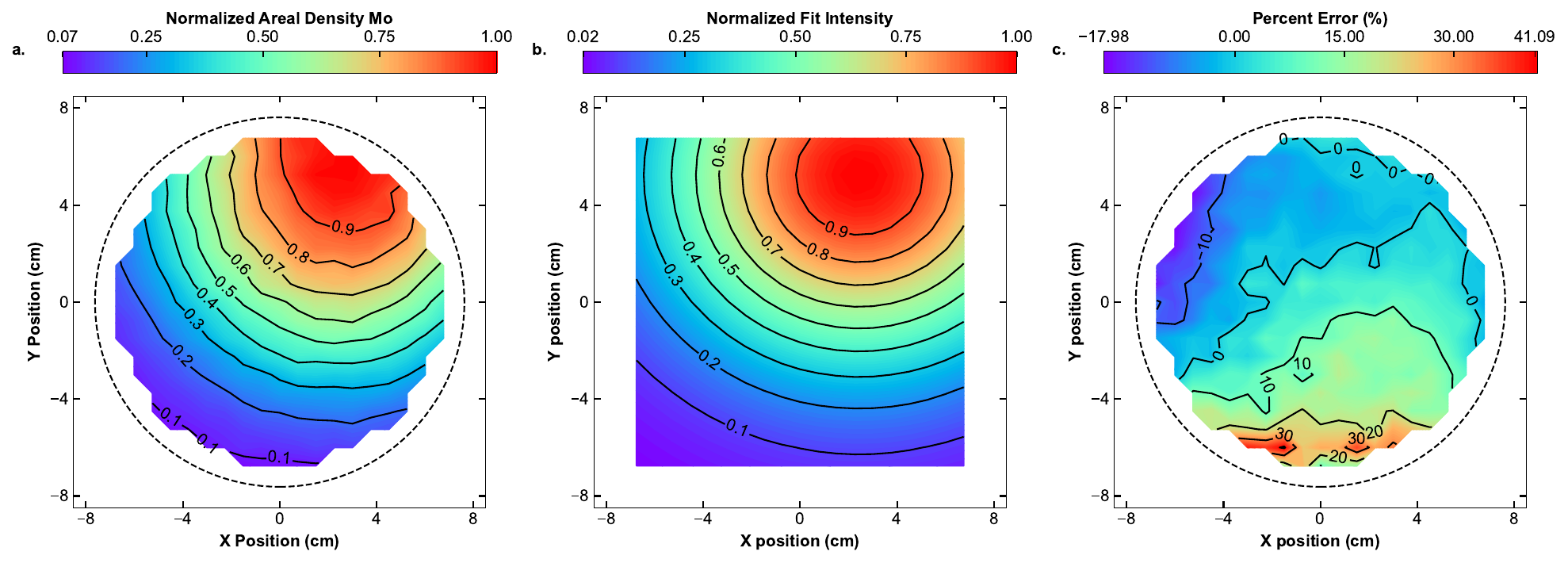}
        \caption{(a) Normalized areal mass density profile of a 2.6~W~cm$^{-2}$ Mo deposition on a 6-inch diameter silicon wafer measured using WDXRF. (b) Two-dimensional Gaussian profile optimized using the areal mass density profile shown in panel (a). (c) Percent error difference between the measured and fit profiles in panels (a) and (b), respectively. In each panel, warm colors correspond to values with higher magnitudes, while cooler colors correspond to values with lower magnitudes, as detailed by the color bars above each respective plot and labeled contour lines. The black dotted circle in panels (a) and (c) represents the extent of the 6-inch diameter Si wafer.}

        \label{fig:single}

    \end{figure*}

    An example WDXRF areal mass density profile from a 2.6~W~cm$^{-2}$ Mo deposition is shown in Fig.~\ref{fig:single}(a), where values have been normalized to the maximum measured across the wafer (2.65~$\times~10^{-4}$~g~cm$^{-2}$). The location of the maximum concentration lies in the corner of the wafer on which the Mo target was focused, which is surrounded by a decaying areal mass density profile. Shown in Fig.~\ref{fig:single}(b) is the two-dimensional Gaussian peak shape (Equation~\ref{eq:comp}) used to fit the areal mass density of this deposition, which is rationalized due to the shape of sputter target race tracks:\cite{mahieu_monte_2006}

    \begin{equation} \label{eq:comp}
         m_i(x, y) = \frac{At}{\sigma_{x}\sigma_{y}2\pi}e^{-\frac{1}{2}\left[\frac{(x-x_{0})^2}{\sigma_{x}^{2}}+\frac{(y-y_{0})^2}{\sigma_{y}^{2}}\right]},
        \end{equation}

    \noindent where $\it{m}$$\rm{_i}$ is the areal mass density of species $\it{i}$ (Mo, in this case) at a point with Cartesian coordinates ($\it{x}$,$\it{y}$), $\it{A}$ is the amplitude of the peak, $t$ is the deposition time, $\sigma$${_x}$ and $\sigma$${_y}$ are the standard deviations of the profile along the Cartesian directions, and $\it{x}$$_0$ and $\it{y}$$_0$ are the Cartesian coordinates of the center position of the Gaussian distribution. Of these, $\it{A}$, $\sigma$${_x}$, $\sigma$${_y}$, $\it{x}$$_0$, and $\it{y}$$_0$ are fitting parameters, which are optimized through least squares fitting. It should be noted that the most practical form of this Gaussian shape for this analysis is obtained with rotation fixed, as discussed in the supplemental information and detailed in Table~S2. The percent error between the optimized Gaussian profile and the WDXRF-measured areal mass density, shown in Fig.~\ref{fig:single}(c), is calculated using Equation~\ref{eq:err}:
    
    \begin{equation} \label{eq:err}
        \epsilon(x,y) = \frac{I_{\rm{exp.}}(x,y)-I_{\rm{fit}}(x,y)}{I_{\rm{fit}}(x,y)},
        \end{equation}
    
    \noindent where $\epsilon$ is the percent error at a point ($\it{x, y}$), and $\it{I}$$_{\rm{exp.}}$ and $\it{I}$$_{\rm{fit}}$ are the measured and fit intensities at the same point, respectively. It is evident that the Gaussian shape provides the best fit of the profile in the vicinity of the maxima, in contrast to the tails, which display errors approaching -10 and 30\% about 8~cm away from the amplitude center. In the center of the wafer, the fit recovers an error of between 3 and 4\%. \startcolor These errors are likely incurred due to asymmetry in the areal density profile from sputter gun tilting that is not captured using the 2D Gaussian shape. \bluecolor In this case, the locations that display the largest errors also possess the smallest areal mass densities ($\sim$7 \% of the maximum value) and so are expected to provide more minor contributions to compositional errors in these regions in multicomponent films. Based on this comparison, however, majority component compositions in films with dilute targeted stoichiometries will be less trustworthy away from profile maxima and wafer center. \stopcolor Using this fit, it is accordingly possible to predict the areal mass density at every analog position across the wafer, with the most accurate values obtained between the position of maximum amplitude and wafer center. In addition, if it is assumed that the \startcolor thickness of the film is well below the attenuation depth (typically several $\mu$m-mm for transition metals) for the utilized radiation, and that the \stopcolor volume density of the film does not significantly vary with thickness in the regimes considered in this measurement, then this \startcolor profile \stopcolor further allows for the calculation of atomic concentration per area at each position using Equation~\ref{eq:conv_atomic}:

    \begin{equation} \label{eq:conv_atomic}
            c_i(x,y) = \frac{m_i(x,y)N_\mathrm{A}}{\rho_i},
            \end{equation}

    \noindent where $c_i(x,y)$ and $\rho_i$ are the number of atoms per area and molar mass of species $i$, respectively, and $N_{\rm{A}}$ is Avogadros number.

\begin{figure}
    
        \includegraphics[width=0.4\textwidth]{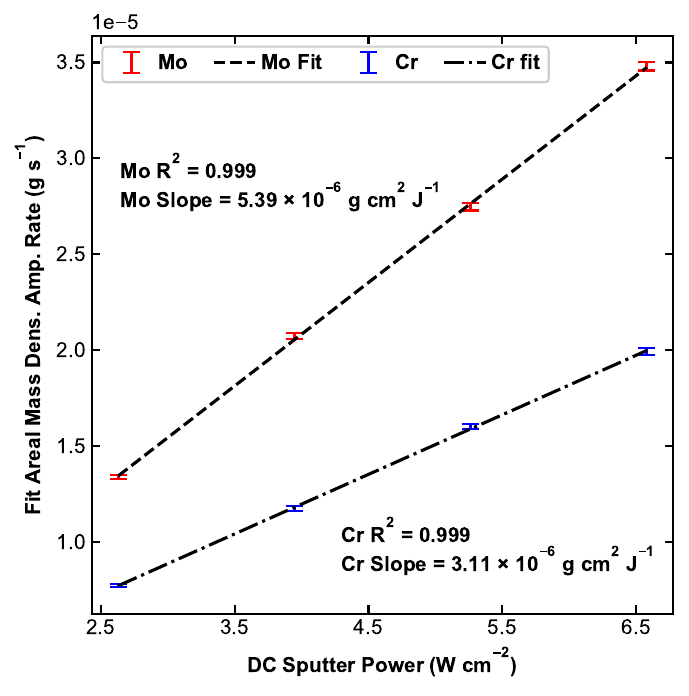}
        \caption{Fit areal mass density amplitude rate of individually deposited Mo (red) and Cr (blue) versus deposition power at $x$ = $x_0$, $y$ = $y_0$. Linear fits to each Mo and Cr data set are shown as black dashed and dot-dashed lines, respectively, the ${R}$$^2$ values and slopes of which are annotated to the left and right. Error bars come from least squares amplitude fitting, and are $\leq$ 1\% for each data point.}

        \label{fig:slopes}

    \end{figure}

    The evolution of this areal mass density profile with applied DC target power was examined by applying an identical fitting procedure to additional films deposited with 3.9, 5.3, and 6.6~W~cm$^{-2}$ applied power densities. Among the five fitting parameters, $\sigma$${_x}$, $\sigma$${_y}$, $\it{x}$$_0$, and $\it{y}$$_0$ are all found to remain essentially constant, displaying averaged values of 5.94~$\pm$~0.03~cm, 5.27~$\pm$~0.12~cm, 5.26~$\pm$~0.06~cm, and 2.37~$\pm$~0.21~cm, respectively, for all investigated sputtering powers (errors represent 95\% confidence intervals). The only parameter of the Gaussian peak shape that varies significantly with power is $\it{A}$, as illustrated in Fig.~\ref{fig:slopes}. The slope of this line (mass per energy) is found to be 5.39~$\times$~$10^{-6}$~g~cm$^2$~J$^{-1}$ when adjusted for target area, areal Mo mass density, and sputter time.
    
    An identical experiment was carried out for Cr. As with Mo, Cr boasts a strongly linear areal mass density amplitude versus sputter power density relationship with a slope of 3.11~$\times$~$10^{-6}$~g~cm$^2$~J$^{-1}$ (also displayed in Fig.~\ref{fig:slopes}) and corresponding nearly constant $\sigma$${_x}$, $\sigma$${_y}$, $\it{x}$$_0$, and $\it{y}$$_0$ parameters (detailed in supplemental information Table S3). Accordingly, \bluecolor assuming negligible effects from target voltage and power drift and chamber conditioning, \stopcolor producing a calculated two-dimensional Gaussian areal mass density profile for a pure metal species in a direct current sputtering procress requires only values for $\sigma$${_x}$, $\sigma$${_y}$, $\it{x}$$_0$, and $\it{y}$$_0$ for a given material, sputter gun, and power supply, which are obtainable through this calibration procedure.

\subsection{Two-Component Deposition Characteristics}

\begin{figure*}
    
        \includegraphics[width=1\textwidth]{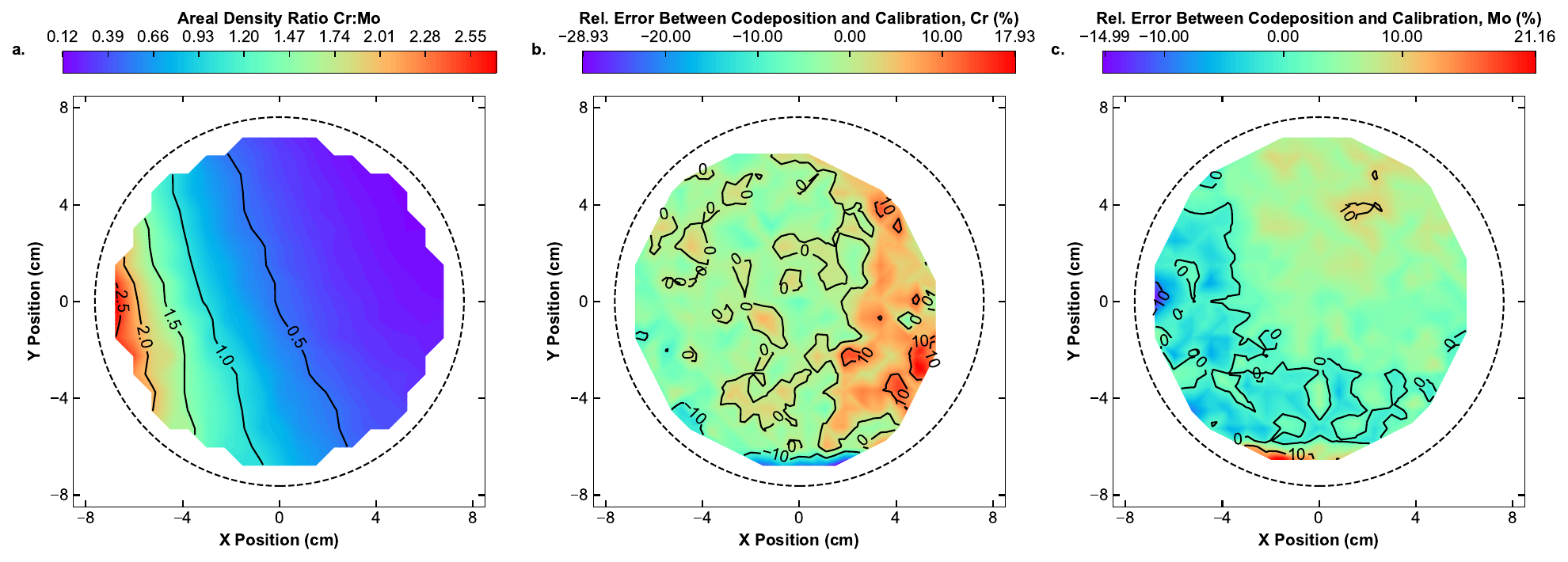}
        \caption{(a) Areal mass density ratio between Cr and Mo of wafer codeposited with 3.9~W~cm$^{-2}$ applied power density across both targets. Percent error between each individual (b) Cr and (c) Mo species in the codeposited wafer and associated 3.9~W~cm$^{-2}$ calibration deposition. In each panel, warm colors correspond to values with higher magnitudes, while cooler colors correspond to values with lower magnitudes, as detailed by the color bars above each respective plot and labeled contour lines. The black dotted circle in all three panels represents the extent of the 6-inch dimater Si wafer in the codeposition.}

        \label{fig:codep}

    \end{figure*}

    Through varying sputter power and monitoring changes in areal mass density profiles, it is evident that single-component depositions can be adequately calculated within a wide power range. It is not immediately obvious based on these two seperate experiments, however, if the same behavior should be expected from two individually well-behaved processes combined. To explore potential interactions during concurrent depositions, a codeposition of Mo and Cr was carried out with identical sputter power densities of 3.9~W~cm$^{-2}$. The areal mass density Cr:Mo ratio at each point across this co-deposited film is shown in Fig.~\ref{fig:codep}a. Closest to the Mo target and in the vicinity of the Mo compositional maximum, this Cr:Mo ratio is at minimum around 0.1, whereas away from the Mo deposition focal point it increases toward 2.5, as expected for a combinatorial process.
    
    Shown in Fig.~\ref{fig:codep}b and Fig.~\ref{fig:codep}c, respectively, are the relative errors between the individual Cr and Mo components of the codeposition and their identical power individual depositions, calculated using an expression similar to Equation~\ref{eq:err}. To make these comparisons direct, the areal mass density profiles from the individual depositions have been shifted in the $\it{x}$ and $\it{y}$-directions to match the Gaussian peak centers of the codeposited wafer to account for translational errors incurred during wafer placement onto the custom deposition stage. When accounting for these shifts, errors generally vary by $\pm$~10\% for both species across the wafer, with amplitude centers displaying errors approaching 0\%. Moreover, save for the variation in $\it{x}$$_0$ and $\it{y}$$_0$ positions due to wafer placement, $\sigma$${_x}$ and $\sigma$${_y}$ parameters fit from the codeposited profiles vary by $\pm$~3\% compared to the individual depositions, which are similar to the errors recovered between individual depositions conducted with different sputter power densities. In addition, Gaussian fits to both individual Cr and Mo areal mass density profiles coexisting within the co-deposited film produce amplitudes that are each $\le$~$\pm$~0.5\% different than their counterparts deposited individually with the same power. This strong agreement indicates that the areal mass density profiles in the codeposited wafers are accurately approximated by the individual component calibration measurements and $\it{vice}$ $\it{versa}$. It is therefore equivalent to calibrate sputter powers for multiple species either individually or simultaneously through codeposition, where the latter partially mitigates error associated with wafer placement.

\subsection{Calibration Process for an Arbitrary Number of Components}

\begin{figure}
    
        \includegraphics[width=0.4\textwidth]{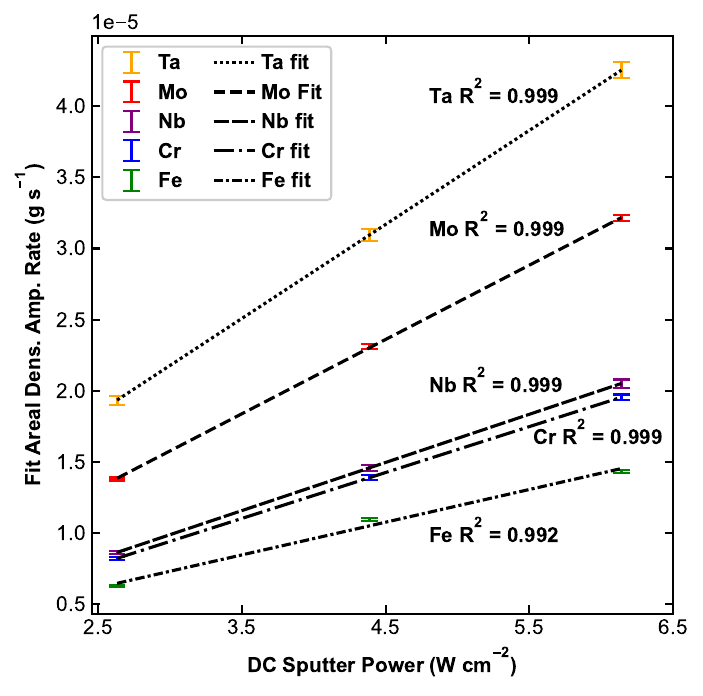}
        \caption{Fit areal mass density amplitude rate of co-deposited Ta (orange), Mo (red), Nb (purple), Cr (blue), and Fe (green) versus deposition power. Linear fits to each data set are shown as black dashed and dotted lines (see legend) and associated $R$$^2$ values are annotated above or below each corresponding line. Error bars come from least squares fitting of amplitude, and are $\leq$~2\% for each data point.}

        \label{fig:five_comp}

    \end{figure}

    Shown in Fig.~\ref{fig:five_comp} are the areal mass density amplitudes of five species, Cr, Fe, Mo, Nb and Ta, co-deposited with each component at three different sputtering powers, 2.63, 4.38, and 6.14~W~cm$^{-2}$. In each case, Gaussian amplitudes are expectedly strongly linear with deposition power. For example, each amplitude linear fit displays an ${R}$$^2$ value $\ge$~0.999 except for Fe. Fitting parameters and associated errors for each species are detailed in supplemental Table S4. As expected, Mo and Cr fitting parameters optimized using the areal mass density profiles from the combinatorial wafers deposited with different uniform powers agree with those from the singular and co-deposited processes. \bluecolor In addition, a drift in supply voltages and currents of 1~$\pm$~2 \% that occurred during each deposition due to target surface erosion, chamber conditioning, and gun temperature equilibration did not result in any observable deviations from expected Gaussian areal mass density profile behavior. \stopcolor Accordingly, a minimum of two depositions at different target powers (ideally above and below the powers expected for the optimized process and away from non-linear power regimes) are required to simultaneously calculate areal density profiles for an arbitrary number of components.

\subsection{Achieving Targeted Compositions}

    Conversion from individual areal mass densities to atomic ratios is straightforward assuming that the film density and thickness do not significantly change within the spot size used for WDXRF measurement. For example, if the film thickness is assumed to be constant within the WDXRF spot, then comparison between atomic areal densities of different species within a single measurement can be considered quantitative because the film volume within the spot is fixed. \startcolor Importantly, use of this assumption enables the analysis of compositions without knowledge of film thickness, making measurements comparable across the wafer. \stopcolor Therefore, at each discrete measurement point, the atomic ratio can be evaluated using Equation~\ref{eq:std}, which computes the root mean squared (RMS) deviation of all species, adjusted by a factor, $C_i$, that takes into account the desired ratio of each species $i$:

\begin{equation} \label{eq:std}
%
        \sigma(x, y) = \sqrt[\leftroot{-2}\uproot{5}\frac{}{}]{\frac{1}{N}\sum_{i = 1}^{N}\left(\frac{c_i(x, y)}{C_i}-\frac{n}{M}\right)^2},
        \end{equation}

    \noindent where $\sigma$($\it{x, y}$) is the RMS deviation at a specified point ($\it{x, y}$) on the wafer, $\it{N}$ is the number of elements considered, $\it{c_i}$ is the measured or simulated atomic areal density of species $\it{i}$ at ($\it{x, y}$), $\it{C_i}$ is the desired atomic ratio of species $\it{i}$ at ($\it{x, y}$) (for example, $\it{C_{\rm{Cr}}}$ = 2 and $\it{C_{\rm{Mo}}}$ = 1 if the desired composition is Cr$_2$Mo), and $n$ and $M$ are the total atomic areal density of all species at ($\it{x, y}$) and the sum of all $C_i$, respectively. This calculation can be completed for every point on a simulated wafer, where the point that produces a minimum value for $\sigma$ displays an atomic composition that is closest to specification.

    As the location of the point with composition closest to that specified can be computed for any simulation, all that remains to determine an optimized power combination is considering every power combination in a brute force manner. Then, a combinatorial deposition can be simulated using each possible power combination, where each contains a point where $\sigma$ is minimized. In practice, it is often most desirable to target a film composition near the wafer center ($\it{x}$~=~$\it{y}$~=~0), which conveniently reduces computing overhead to perform this optimization while also focusing on a wafer region where the total measurement error for all species is most evenly distributed. Even so, each calibration shown in Fig.~\ref{fig:five_comp} spans 40 integer powers, which leads to 1~$\times$~$10^{8}$ possible permutations, where reduction to four or three components produces 2.5~$\times$~$10^{6}$ and 6.4~$\times$~$10^{4}$ possible permutations, respectively. In cases where the number of components produces a prohibitive computing overhead, an analytical solution may prove more useful.

    Upon inspection of Equation~\ref{eq:std}, it is evident that an exact minimum in RMS deviation is obtained if each species displays a composition that fulfills Equation~\ref{eq:perf} at ($\it{x, y}$):

\begin{equation} \label{eq:perf}
        c_i(x,y) = \frac{nC_i}{M},
        \end{equation}

    Further, by including an adjustment for stoichiometry, Equation~\ref{eq:comp} can be simplified to describe only the deposition rate as a function of applied target power at a specified position ($\it{x}$, $\it{y}$), as related in Equation~\ref{eq:comp_simple}:

\begin{equation} \label{eq:comp_simple}
                 c_i(P,x,y) = \frac{N_\mathrm{A}\alpha_{i}(x,y)}{\rho_{i}C_{i}}[\mathcal{A}_{i}P_i+b_i],
                \end{equation}
    
    \noindent where $\mathcal{A}_{i}$ and $b_i$ are the areal mass density amplitude slope and intercept, respectively, of species $i$, $P_i$ is the input power, $\alpha_{i} (x,y)$ is a prefactor that contains the non-amplitude components of the Gaussian profile, and $N_{\rm{A}}$ and $\rho_i$ are Avogadros number and molar mass, respectively (necessary to include because $\mathcal{A}_{i}$ and $b_i$ are in units of mass instead of atoms).
    
    Combining Equations~\ref{eq:perf} and \ref{eq:comp_simple} and rearranging gives Equation~\ref{eq:power}:

\begin{equation} \label{eq:power}
                 P_i = \mathcal{A}_i^{-1}\left(\frac{\rho_inC_i}{N_\mathrm{A}\alpha_i(x,y)M}-b_i\right),
                \end{equation}   
    
    \noindent which relates the power required to obtain the targeted atomic ratios ($C_i$) and total atomic density ($n$) to the deposition parameters obtained through calibration. The generalized form of Equation~\ref{eq:power}, which does not assume the non-covariance established in Sec.IIIB and IIIC, is given as supplemental information Equation S1.
    
    If a targeted total atomic density of 6~$\times$~10$^{18}$~at.~cm$^{-2}$ is input (the approximate per-species areal atomic density at $x$~=~$y$~=~0 for the 2.63~W~cm$^2$ five-component calibration), then using the calibration parameters given in supplemental Table S4 produces power densities of 3.10, 3.52, 2.84, 4.44, and 3.13 W~cm$^{-2}$ for Cr, Fe, Mo, Nb, and Ta targets, respectively. For the target supplies utilized for these depositions, integer inputs for power (in W) are required, which necessitates the rounding of each optimized power density, and likely induces positional and composition errors.
    
    Shown in Fig.~\ref{fig:combi_dep_sigma}(a) is a map of the relative $\sigma$ of a combinatorial deposition simulated using rounded optimized powers of 35, 40, 32, 50, and 36 W for Cr, Fe, Mo, Nb, and Ta targets, respectively. The minimum relative $\sigma$ present in this simulated deposition resides at the wafer center, and is surrounded by a radially increasing profile with maxima at the grid corners. Shown in Fig.~\ref{fig:combi_dep_sigma}(b) is the relative $\sigma$ map for the corresponding combinatorial deposition, which displays strong agreement with simulation. The white star in Fig.~\ref{fig:combi_dep_sigma}(b) is the location of the measurement that reports the stoichiometry that is closest to equiatomic. This point, located at $x$ = 0.75 cm, $y$ = 0 cm, displays an atomic composition of Cr$_{0.19}$Fe$_{0.20}$Mo$_{0.19}$Nb$_{0.21}$Ta$_{0.21}$ according to WDXRF. Deviations in this location from exact center likely arise due to wafer placement into the deposition system as well as from the rounding of optimized power values. Seperately, deviations from a perfect equiatomic composition are contributed to by target power rounding, translation resolution, and the error accumulated by each species away from peak WDXRF intensity amplitudes. Relative abundances for each element within the combinatorial film are shown in supplemental Fig.~S1. For all species, abundances decay away from target foci, with atomic ratio regions of 0.2 intersecting near the wafer center. Maximum ratios for each species are around 0.45 - 0.55, and minima all fall near 0.05. In total, the agreement between simulation and deposition validates this optimization procedure, which allows for the facile synthesis of combinatorial wafers with targeted compositions.

    \begin{figure}
    
        \includegraphics[width=0.4\textwidth]{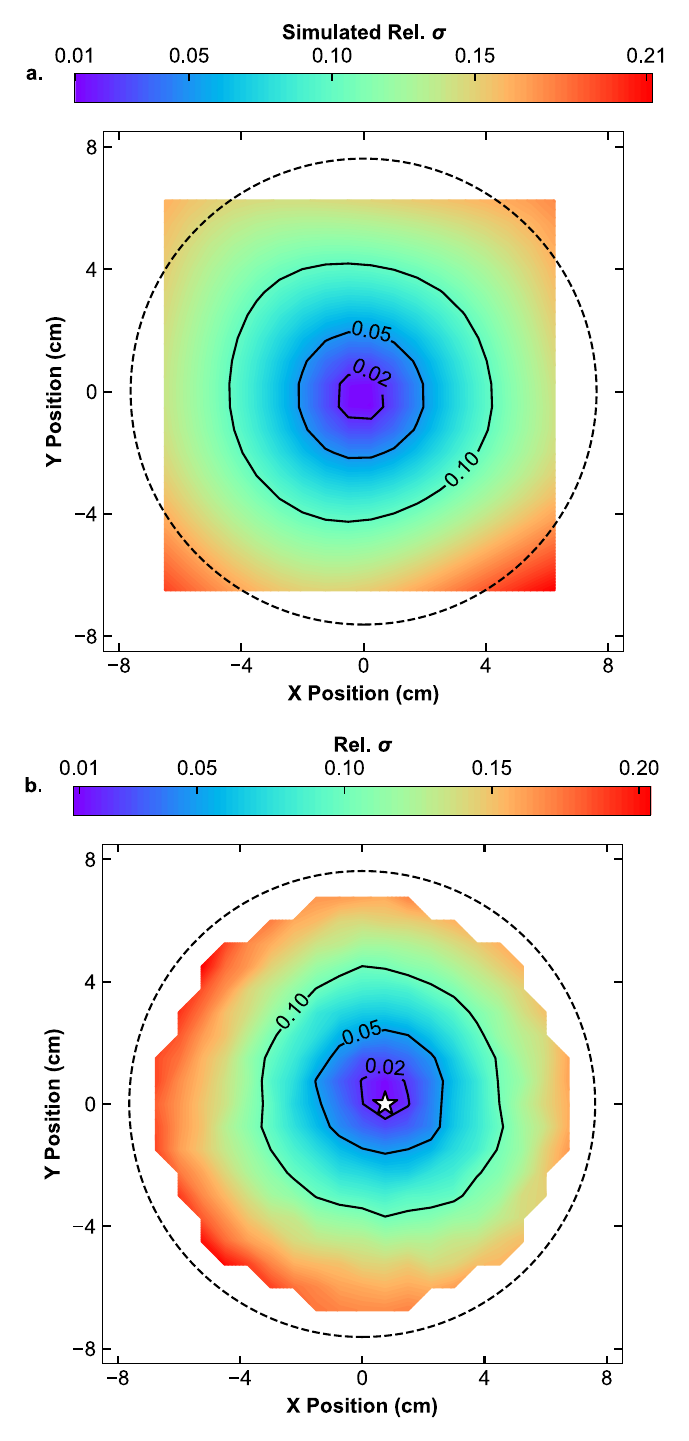}
        \caption{Relative RMS deviation maps for a (a) simulated and (b) deposited Cr$_{v}$Fe$_{w}$Mo$_{x}$Nb$_{y}$Ta$_{z}$ film synthesized using powers optimized to obtain an equiatomic composition at the center position. The white star near the center of the map in panel (b) corresponds to the position with the lowest relative $\sigma$ value. In all panels, warm and cool colors indicate larger and smaller values, respectively, as detailed by each corresponding color bar. The black dotted circle in both panels represents the extent of the 6-inch dimater Si wafer in the codeposition.}

        \label{fig:combi_dep_sigma}

    \end{figure}

\subsection{Cross-Technique Composition Verification}

    For cross-technique comparison, EDS spectra were collected at six locations across the combinatorial wafer, each with a corresponding WDXRF measurement. A summary of the comparison is given in Table~\ref{tab:comp_compare}, whereas a direct comparison of compositions at each point is provided in supplemental Fig.~S2, with the locations of each annotated in supplemental Fig.~S1. Comparing the composition measurements between the techniques produces an average difference of $\pm$~2.8~at.~\%. These differences are concentrated mainly Fe and Nb, which display average differences of 5.0 and 4.9~at.~\%, respectively, in contrast to Cr, Mo, and Ta, which all maintain average differences of \textless~1.6\%. In particular, across all points EDS uniformly measures Fe and Nb compositions that are less than and greater than WDXRF, respectively, which indicates that the use of a singular standard may be required to obtain more accurate agreement between these two techniques, and would likely improve quantitative accuracy of both. \startcolor Regardless, the agreement between these spot measurements validates the selection of WDXRF spot size and overlap, which was otherwise selected to maximize point density, sample signal, and throughput.\stopcolor

    \begin{table}
        \caption{\label{tab:comp_compare}Average absolute percent difference between WDXRF and EDS measurements and associated standard deviations for each element across six locations.}
        \begin{ruledtabular}
        \begin{tabular}{ccc}
        Species&Av. Abs. Difference (at. \%)&Std. Dev. (at. \%)\\
        \hline
        Cr&1.4&0.6\\
        Fe&5.0&1.6\\
        Mo&1.6&0.7\\
        Nb&4.9&2.1\\
        Ta&1.1&0.7\\
        Total&2.8&0.7\\
        \end{tabular}
        \end{ruledtabular}
    \end{table}

\section{Discussion}

    Based upon the strongly linear power density versus amplitude relationships shown in Fig.~\ref{fig:five_comp}, this methodology posesses several capabilities that are not immediately displayed by a single demonstration. Primarily, following a minimum of two calibration depositions, it is evident that it is possible to obtain conditions for a wide range of center compositions, which is enabling for the high-throughput investigation of full stoichiometric spaces of material systems. In addition, due to the strongly linear relationship between all species and applied target power, adjustment of powers by a scalar is expected to produce similar compositions (assuming a negligable effect of areal mass density amplitude intercept on Equation~\ref{eq:power}), meaning that calibration is capable of suggesting multiple growth rates. An example comparison of depositions simulated using the previously optimized power densities scaled by different values is shown in supplemental information Fig.~S3, which supports this behavior. \bluecolor Asymmetric peaks shapes may also be utilized, instead of a fixed-rotation Gaussian, to accommodate more tilted gun geometries. In these cases, it should be ensured that areal mass density profiles do not geometrically change with variations in power, or that such changes are captured by calibration. \stopcolor Further, calibration can be completed using EDS maps or point measurements (instead of WDXRF) if film thicknesses are considered that are well below the interaction depth typical of alloys ($\sim$1-2 $\mu$m), and customized equipment (i.e. the sputter gun flange or stage employed within this study) is not required for combinatorial sputtering assuming the employed stage is capable of height translation. \startcolor This optimized process is then capable of producing films with thicknesses that are greater than the limit imposed by the attenuation depth of the chosen composition measurement technique. \stopcolor In addition, wafer size and material (assuming other utilized substrates are similarly smooth) are not expected to strongly affect the composition and deposition rates of \startcolor films that are sufficiently thick to achieve non-heterogeneous growth (i.e. ~10 nm), \stopcolor meaning that calibration can be completed using readily-available Si before other, more valuable substrates are used for subsequent experimentation. For example, an analysis of a combinatorial film deposited on an 8 inch-diameter Si wafer using the same optimized power densities as the deposition analyzed in Fig.~\ref{fig:combi_dep_sigma}, shown in supplemental information Fig.~S4, displays an identically centered $\sigma$ minimum.

    Additionally, radio frequency (rf) sputtering has also been shown to display strongly linear power versus growth rate characteristics,\cite{thelen_python-based_2025} and can therefore be utilized for species that display excessively high DC growth rates, such as Cu or Sb. For materials that do not sputter quickly, this method is also capable of accomodating multiple targets of the same species. Further, the strong agreement between individual and codeposition indicates that a new species can be calibrated alone and incorporated into an already-optimized process, and that an already-optimized process remains functional even if one or more targets is removed.

    Given the asymmetric conditions required for combinatorial synthesis, this methodology also posesses several limitations. For example, deposition characteristics are strongly dependent on wafer height and gun tilt, and changes of either would be expected to nullify any calibration progress using this method. \bluecolor Similarly, a new background pressure during sputtering would be expected to disproportionately affect the Gaussian areal mass density profile shapes of different target species and thereby requires a new calibration. \stopcolor In parallel, changes in wafer temperature, which may occur during a deposition with large numbers of components, large powers, and long growth times, may reduce comparability between depositions conducted at different rates or for significantly different durations. Such variation may be mitigated through use of a cooled stage. The application of this procedure is also limited in predicting film compositions in reactive sputtering processes due to the different reactive potentials of each alloy species,\cite{fields_high-throughput_2025} although it may provide intuition for the distribution of the metallic elements. \bluecolor Finally, periodic `spot-checks' or recalibrations of a calibrated process may be necessary to account for drift in background pressures and power supplies and changes in target erosion profile and chamber conditioning, although it should be noted that such variations were not observed throughout the depositions described herein.\stopcolor

\section{Conclusion}

    A methodology to calibrate combinatorial sputtering processes for targeted compositions has been introduced that relies on the strongly linear relationship between target power and deposition rate and non-covariant relationships between the deposition rates of multiple sources. Using this procedure, it is possible to create a model for an arbitrary number of components that is capable of suggesting conditions to obtain targeted compositions at specific wafer coordinates. The production of this model requires at minimum two calibration depositions at target power extremes, and has been demonstrated on a CrFeMoNbTa alloy combinatorial wafer having an equiatomic position near wafer center, as verified by both WDXRF and EDS. This process is enabling for the high-throughput preparation of metallic combinatorial libraries, where deposition rate optimization often leads to synthesis bottlenecks.

    \section{Supplementary Material}

    The supplementary material contains a discussion of the drawbacks of rotatable 2D Gaussian peak shapes, tables containing optimized 2D Gaussian parameters, a generalized optimization expression, additional WDXRF composition measurements and simulations, and a comparison between WDXRF mapped compositions and EDS point measurements.

    \section{Acknowledgement}

    This work was supported by the Office of Naval Research through the Naval Research Laboratory's Basic Research Program.

\section{Author Declarations}

    The authors have no conflicts to disclose.

\section{Data Availability Statement}

    The data that support the findings of this study are available from the corresponding author upon reasonable request.

\bibliographystyle{unsrt}

\bibliography{bibliography}

@article{iwasaki_machine_2021,
	title = {Machine learning autonomous identification of magnetic alloys beyond the {Slater}-{Pauling} limit},
	volume = {2},
	issn = {2662-4443},
	url = {https://www.nature.com/articles/s43246-021-00135-0},
	doi = {10.1038/s43246-021-00135-0},
	language = {en},
	number = {1},
	urldate = {2023-11-09},
	journal = {Commun. Mater.},
	author = {Iwasaki, Yuma and Sawada, Ryohto and Saitoh, Eiji and Ishida, Masahiko},
	month = mar,
	year = {2021},
	note = {Number: 1},
	pages = {31},
}

@article{green_applications_2013,
	title = {Applications of high throughput (combinatorial) methodologies to electronic, magnetic, optical, and energy-related materials},
	volume = {113},
	issn = {0021-8979, 1089-7550},
	url = {https://pubs.aip.org/jap/article/113/23/231101/376956/Applications-of-high-throughput-combinatorial},
	doi = {10.1063/1.4803530},
	language = {en},
	number = {23},
	urldate = {2024-07-15},
	journal = {J. Appl. Phys.},
	author = {Green, Martin L. and Takeuchi, Ichiro and Hattrick-Simpers, Jason R.},
	month = jun,
	year = {2013},
	note = {Number: 23},
	pages = {231101},
}

@article{siol_combinatorial_2016,
	title = {Combinatorial {Reactive} {Sputtering} of {In}$_{\textrm{2}}${S}$_{\textrm{3}}$ as an {Alternative} {Contact} {Layer} for {Thin} {Film} {Solar} {Cells}},
	volume = {8},
	issn = {1944-8244, 1944-8252},
	url = {https://pubs.acs.org/doi/10.1021/acsami.6b02213},
	doi = {10.1021/acsami.6b02213},
	language = {en},
	number = {22},
	urldate = {2024-07-15},
	journal = {ACS Appl. Mater. Inter.},
	author = {Siol, Sebastian and Dhakal, Tara P. and Gudavalli, Ganesh S. and Rajbhandari, Pravakar P. and DeHart, Clay and Baranowski, Lauryn L. and Zakutayev, Andriy},
	month = jun,
	year = {2016},
	note = {Number: 22},
	pages = {14004--14011},
}

@article{shi_high-throughput_2020,
	title = {High-throughput synthesis and corrosion behavior of sputter-deposited nanocrystalline {Al} ({CoCrFeNi})100-combinatorial high-entropy alloys},
	volume = {195},
	issn = {02641275},
	url = {https://linkinghub.elsevier.com/retrieve/pii/S0264127520305530},
	doi = {10.1016/j.matdes.2020.109018},
	language = {en},
	urldate = {2024-07-15},
	journal = {Mater. Des.},
	author = {Shi, Yunzhu and Yang, Bin and Rack, Philip D. and Guo, Shifeng and Liaw, Peter K. and Zhao, Ying},
	month = oct,
	year = {2020},
	pages = {109018},
}

@article{kauffmann_combinatorial_2017,
	title = {Combinatorial exploration of the {High} {Entropy} {Alloy} {System} {Co}-{Cr}-{Fe}-{Mn}-{Ni}},
	volume = {325},
	issn = {02578972},
	url = {https://linkinghub.elsevier.com/retrieve/pii/S0257897217306448},
	doi = {10.1016/j.surfcoat.2017.06.041},
	language = {en},
	urldate = {2024-07-19},
	journal = {Surf. Coat. Technol.},
	author = {Kauffmann, Alexander and Stüber, Michael and Leiste, Harald and Ulrich, Sven and Schlabach, Sabine and Szabó, Dorothée Vinga and Seils, Sascha and Gorr, Bronislava and Chen, Hans and Seifert, Hans-Jürgen and Heilmaier, Martin},
	month = sep,
	year = {2017},
	pages = {174--180},
}

@article{fackler_combinatorial_2017,
	title = {Combinatorial study of {Fe}-{Co}-{V} hard magnetic thin films},
	volume = {18},
	issn = {1468-6996, 1878-5514},
	url = {https://www.tandfonline.com/doi/full/10.1080/14686996.2017.1287520},
	doi = {10.1080/14686996.2017.1287520},
	language = {en},
	number = {1},
	urldate = {2024-07-19},
	journal = {Sci. Technol. Adv. Mat.},
	author = {Fackler, Sean W. and Alexandrakis, Vasileios and König, Dennis and Kusne, A. Gilad and Gao, Tieren and Kramer, Matthew J. and Stasak, Drew and Lopez, Kenny and Zayac, Brad and Mehta, Apurva and Ludwig, Alfred and Takeuchi, Ichiro},
	month = dec,
	year = {2017},
	note = {Number: 1},
	pages = {231--238},
}

@article{fields_high-throughput_2025,
	title = {High-throughput material search by magnetic and compositional mapping of reactively sputtered combinatorial {FexVyNz} films},
	volume = {137},
	issn = {0021-8979, 1089-7550},
	url = {https://pubs.aip.org/jap/article/137/2/025304/3330266/High-throughput-material-search-by-magnetic-and},
	doi = {10.1063/5.0243499},
	language = {en},
	number = {2},
	urldate = {2025-01-13},
	journal = {J. Appl. Phys.},
	author = {Fields, Shelby S. and Van ‘T Erve, Olaf M. J. and McGrath, Andrew and Johnson, Francis and Bennett, Steven P.},
	month = jan,
	year = {2025},
	pages = {025304},
}

@article{snyder_material_2015,
	title = {Material optimization via combinatorial deposition and analysis for thermoelectric thin films},
	volume = {596},
	issn = {00406090},
	url = {https://linkinghub.elsevier.com/retrieve/pii/S0040609015008305},
	doi = {10.1016/j.tsf.2015.08.054},
	language = {en},
	urldate = {2025-06-06},
	journal = {Thin Solid Films},
	author = {Snyder, Ryan D. and Thomas, Evan L. and Voevodin, Andrey A.},
	month = dec,
	year = {2015},
	pages = {233--241},
}

@article{mcginn_thin-film_2019,
	title = {Thin-{Film} {Processing} {Routes} for {Combinatorial} {Materials} {Investigations}—{A} {Review}},
	volume = {21},
	copyright = {https://doi.org/10.15223/policy-029},
	issn = {2156-8952, 2156-8944},
	url = {https://pubs.acs.org/doi/10.1021/acscombsci.9b00032},
	doi = {10.1021/acscombsci.9b00032},
	language = {en},
	number = {7},
	urldate = {2025-06-06},
	journal = {ACS Comb. Sci.},
	author = {McGinn, Paul J.},
	month = jul,
	year = {2019},
	pages = {501--515},
}

@article{choi_combinatorial_2015,
	title = {Combinatorial {Search} for {High}‐{Activity} {Hydrogen} {Catalysts} {Based} on {Transition}‐{Metal}‐{Embedded} {Graphitic} {Carbons}},
	volume = {5},
	copyright = {http://onlinelibrary.wiley.com/termsAndConditions\#vor},
	issn = {1614-6832, 1614-6840},
	url = {https://onlinelibrary.wiley.com/doi/10.1002/aenm.201501423},
	doi = {10.1002/aenm.201501423},
	language = {en},
	number = {23},
	urldate = {2025-06-06},
	journal = {Adv. Energy Mater.},
	author = {Choi, Woon Ih and Wood, Brandon C. and Schwegler, Eric and Ogitsu, Tadashi},
	month = dec,
	year = {2015},
	pages = {1501423},
}

@article{jin_combinatorial_2013,
	title = {Combinatorial search of superconductivity in {Fe}-{B} composition spreads},
	volume = {1},
	issn = {2166-532X},
	url = {https://pubs.aip.org/apm/article/1/4/042101/120026/Combinatorial-search-of-superconductivity-in-Fe-B},
	doi = {10.1063/1.4822435},
	language = {en},
	number = {4},
	urldate = {2025-06-06},
	journal = {APL Mater.},
	author = {Jin, Kui and Suchoski, Richard and Fackler, Sean and Zhang, Yi and Pan, Xiaoqing and Greene, Richard L. and Takeuchi, Ichiro},
	month = oct,
	year = {2013},
	pages = {042101},
}

@article{garcia_combinatorial_2007,
	title = {Combinatorial {Synthesis} and {Hydrogenation} of {Mg}/{Al} {Libraries} {Prepared} by {Electron} {Beam} {Physical} {Vapor} {Deposition}},
	volume = {9},
	issn = {1520-4766, 1520-4774},
	url = {https://pubs.acs.org/doi/10.1021/cc060131h},
	doi = {10.1021/cc060131h},
	language = {en},
	number = {2},
	urldate = {2025-06-06},
	journal = {J. Comb. Chem.},
	author = {Garcia, Gemma and Doménech-Ferrer, Roger and Pi, Francesc and Santiso, Josep and Rodríguez-Viejo, Javier},
	month = mar,
	year = {2007},
	pages = {230--236},
}

@article{logvenov_combinatorial_2007,
	title = {Combinatorial molecular beam epitaxy of {La2}-{xSrxCuO4}+y},
	volume = {460-462},
	copyright = {https://www.elsevier.com/tdm/userlicense/1.0/},
	issn = {09214534},
	url = {https://linkinghub.elsevier.com/retrieve/pii/S0921453407002365},
	doi = {10.1016/j.physc.2007.03.408},
	language = {en},
	urldate = {2025-06-06},
	journal = {Physica C},
	author = {Logvenov, G. and Sveklo, I. and Bozovic, I.},
	month = sep,
	year = {2007},
	pages = {416--419},
}

@article{schenck_high_2004,
	title = {High throughput characterization of the optical properties of compositionally graded combinatorial films},
	volume = {223},
	copyright = {https://www.elsevier.com/tdm/userlicense/1.0/},
	issn = {01694332},
	url = {https://linkinghub.elsevier.com/retrieve/pii/S0169433203009206},
	doi = {10.1016/j.apsusc.2003.07.005},
	language = {en},
	number = {1-3},
	urldate = {2025-06-06},
	journal = {Appl. Surf. Sci.},
	author = {Schenck, Peter K and Kaiser, Debra L and Davydov, Albert V},
	month = feb,
	year = {2004},
	pages = {200--205},
}

@article{li_high-temperature_2019,
	title = {High-temperature bulk metallic glasses developed by combinatorial methods},
	volume = {569},
	issn = {0028-0836, 1476-4687},
	url = {https://www.nature.com/articles/s41586-019-1145-z},
	doi = {10.1038/s41586-019-1145-z},
	language = {en},
	number = {7754},
	urldate = {2025-06-06},
	journal = {Nature},
	author = {Li, Ming-Xing and Zhao, Shao-Fan and Lu, Zhen and Hirata, Akihiko and Wen, Ping and Bai, Hai-Yang and Chen, MingWei and Schroers, Jan and Liu, YanHui and Wang, Wei-Hua},
	month = may,
	year = {2019},
	pages = {99--103},
}

@article{sur_high_2023,
	title = {A {High} {Throughput} {Aqueous} {Passivation} {Testing} {Methodology} for {Compositionally} {Complex} {Alloys} {Using} a {Scanning} {Droplet} {Cell}},
	volume = {170},
	issn = {0013-4651, 1945-7111},
	url = {https://iopscience.iop.org/article/10.1149/1945-7111/aceeb8},
	doi = {10.1149/1945-7111/aceeb8},
	language = {en},
	number = {8},
	urldate = {2025-06-06},
	journal = {J. Electrochem. Soc.},
	author = {Sur, Debashish and Joress, Howie and Hattrick-Simpers, Jason and Scully, John R.},
	month = aug,
	year = {2023},
	pages = {081507},
}

@article{suram_combinatorial_2015,
	title = {Combinatorial thin film composition mapping using three dimensional deposition profiles},
	volume = {86},
	issn = {0034-6748, 1089-7623},
	url = {https://pubs.aip.org/rsi/article/86/3/033904/361474/Combinatorial-thin-film-composition-mapping-using},
	doi = {10.1063/1.4914466},
	language = {en},
	number = {3},
	urldate = {2025-06-06},
	journal = {Rev. Sci. Instrum.},
	author = {Suram, Santosh K. and Zhou, Lan and Becerra-Stasiewicz, Natalie and Kan, Kevin and Jones, Ryan J. R. and Kendrick, Brian M. and Gregoire, John M.},
	month = mar,
	year = {2015},
	pages = {033904},
}

@article{otani_high-throughput_2007,
	title = {A high-throughput thermoelectric power-factor screening tool for rapid construction of thermoelectric property diagrams},
	volume = {91},
	issn = {0003-6951, 1077-3118},
	url = {https://pubs.aip.org/apl/article/91/13/132102/333831/A-high-throughput-thermoelectric-power-factor},
	doi = {10.1063/1.2789289},
	language = {en},
	number = {13},
	urldate = {2025-06-06},
	journal = {Appl. Phys. Lett.},
	author = {Otani, M. and Lowhorn, N. D. and Schenck, P. K. and Wong-Ng, W. and Green, M. L. and Itaka, K. and Koinuma, H.},
	month = sep,
	year = {2007},
	pages = {132102},
}

@article{liu_combinatorial_2022,
	title = {Combinatorial {High}-{Throughput} {Methods} for {Designing} {Hydrogen} {Evolution} {Reaction} {Catalysts}},
	volume = {12},
	copyright = {https://doi.org/10.15223/policy-029},
	issn = {2155-5435, 2155-5435},
	url = {https://pubs.acs.org/doi/10.1021/acscatal.2c00869},
	doi = {10.1021/acscatal.2c00869},
	language = {en},
	number = {7},
	urldate = {2025-06-06},
	journal = {ACS Catal.},
	author = {Liu, Xuanzhi and Zou, Peng and Song, Lijian and Zang, Bowen and Yao, Bingnan and Xu, Wei and Li, Fushan and Schroers, Jan and Huo, Juntao and Wang, Jun-Qiang},
	month = apr,
	year = {2022},
	pages = {3789--3796},
}

@article{marshal_combinatorial_2017,
	title = {Combinatorial synthesis of high entropy alloys: {Introduction} of a novel, single phase, body-centered-cubic {FeMnCoCrAl} solid solution},
	volume = {691},
	issn = {09258388},
	shorttitle = {Combinatorial synthesis of high entropy alloys},
	url = {https://linkinghub.elsevier.com/retrieve/pii/S0925838816327281},
	doi = {10.1016/j.jallcom.2016.08.326},
	language = {en},
	urldate = {2025-06-06},
	journal = {J. Alloy Compd.},
	author = {Marshal, A. and Pradeep, K.G. and Music, D. and Zaefferer, S. and De, P.S. and Schneider, J.M.},
	month = jan,
	year = {2017},
	pages = {683--689},
}

@article{nadaud_enhancement_2024,
	title = {Enhancement of {Piezoelectric} {Properties} in a {Narrow} {Cerium} {Doping} {Range} of {Ba}$_{\textrm{1– \textit{x}}}${Ca}$_{\textrm{\textit{x}}}${Ti}$_{\textrm{1–\textit{y}}}${Zr}$_{\textrm{\textit{y}}}${O}$_{\textrm{3}}$ {Evidenced} by {Combinatorial} {Experiment}},
	volume = {6},
	copyright = {https://doi.org/10.15223/policy-029},
	issn = {2637-6113, 2637-6113},
	url = {https://pubs.acs.org/doi/10.1021/acsaelm.4c01282},
	doi = {10.1021/acsaelm.4c01282},
	language = {en},
	number = {10},
	urldate = {2025-06-06},
	journal = {ACS Appl. Electron. Mater.},
	author = {Nadaud, Kevin and Nataf, Guillaume F. and Jaber, Nazir and Negulescu, Béatrice and Giovannelli, Fabien and Andreazza, Pascal and Birnal, Pierre and Wolfman, Jérôme},
	month = oct,
	year = {2024},
	pages = {7392--7401},
}

@article{thelen_python-based_2025,
	title = {A python-based approach to sputter deposition simulations in combinatorial materials science},
	volume = {503},
	issn = {02578972},
	url = {https://linkinghub.elsevier.com/retrieve/pii/S0257897225002725},
	doi = {10.1016/j.surfcoat.2025.131998},
	language = {en},
	urldate = {2025-06-10},
	journal = {Surf. Coat. Technol.},
	author = {Thelen, F. and Zehl, R. and Bürgel, J.L. and Depla, D. and Ludwig, A.},
	month = may,
	year = {2025},
	pages = {131998},
}

@article{mahieu_monte_2006,
	title = {Monte {Carlo} simulation of the transport of atoms in {DC} magnetron sputtering},
	volume = {243},
	copyright = {https://www.elsevier.com/tdm/userlicense/1.0/},
	issn = {0168583X},
	url = {https://linkinghub.elsevier.com/retrieve/pii/S0168583X05017544},
	doi = {10.1016/j.nimb.2005.09.018},
	language = {en},
	number = {2},
	urldate = {2025-06-10},
	journal = {Nucl. Instrum. Meth. B},
	author = {Mahieu, S. and Buyle, G. and Depla, D. and Heirwegh, S. and Ghekiere, P. and De Gryse, R.},
	month = feb,
	year = {2006},
	pages = {313--319},
}

@article{sasaki_identifying_2020,
	title = {Identifying {Optimal} {Strain} in {Bismuth} {Telluride} {Thermoelectric} {Film} by {Combinatorial} {Gradient} {Thermal} {Annealing} and {Machine} {Learning}},
	volume = {22},
	copyright = {https://doi.org/10.15223/policy-029},
	issn = {2156-8952, 2156-8944},
	url = {https://pubs.acs.org/doi/10.1021/acscombsci.0c00112},
	doi = {10.1021/acscombsci.0c00112},
	language = {en},
	number = {12},
	urldate = {2025-06-12},
	journal = {ACS Comb. Sci.},
	author = {Sasaki, Michiko and Ju, Shenghong and Xu, Yibin and Shiomi, Junichiro and Goto, Masahiro},
	month = dec,
	year = {2020},
	pages = {782--790},
}

@article{al_hasan_combinatorial_2020,
	title = {Combinatorial {Exploration} and {Mapping} of {Phase} {Transformation} in a {Ni}–{Ti}–{Co} {Thin} {Film} {Library}},
	volume = {22},
	copyright = {https://doi.org/10.15223/policy-029},
	issn = {2156-8952, 2156-8944},
	url = {https://pubs.acs.org/doi/10.1021/acscombsci.0c00097},
	doi = {10.1021/acscombsci.0c00097},
	language = {en},
	number = {11},
	urldate = {2025-06-12},
	journal = {ACS Comb. Sci.},
	author = {Al Hasan, Naila M. and Hou, Huilong and Gao, Tieren and Counsell, Jonathan and Sarker, Suchismita and Thienhaus, Sigurd and Walton, Edward and Decker, Peer and Mehta, Apurva and Ludwig, Alfred and Takeuchi, Ichiro},
	month = nov,
	year = {2020},
	pages = {641--648},
}

@article{gebhardt_combinatorial_2012,
	title = {Combinatorial thin film materials science: {From} alloy discovery and optimization to alloy design},
	volume = {520},
	copyright = {https://www.elsevier.com/tdm/userlicense/1.0/},
	issn = {00406090},
	shorttitle = {Combinatorial thin film materials science},
	url = {https://linkinghub.elsevier.com/retrieve/pii/S0040609012005238},
	doi = {10.1016/j.tsf.2012.04.062},
	language = {en},
	number = {17},
	urldate = {2025-06-12},
	journal = {Thin Solid Films},
	author = {Gebhardt, Thomas and Music, Denis and Takahashi, Tetsuya and Schneider, Jochen M.},
	month = jun,
	year = {2012},
	pages = {5491--5499},
}

@article{nagy_mapping_2024,
	title = {Mapping the microstructure and the mechanical performance of a combinatorial {Co}–{Cr}–{Cu}–{Fe}–{Ni}–{Zn} high-entropy alloy thin film processed by magnetron sputtering technique},
	volume = {31},
	issn = {22387854},
	url = {https://linkinghub.elsevier.com/retrieve/pii/S223878542401370X},
	doi = {10.1016/j.jmrt.2024.06.059},
	language = {en},
	urldate = {2025-06-12},
	journal = {J. Mater. Res. Technol.},
	author = {Nagy, Peter and Watroba, Maria and Hegedus, Zoltan and Michler, Johann and Petho, Laszlo and Schwiedrzik, Jakob and Czigany, Zsolt and Gubicza, Jeno},
	month = jul,
	year = {2024},
	pages = {47--61},
}

@article{zakay_combinatorial_2023,
	title = {A {Combinatorial} {Approach} for the {Solution} {Deposition} of {Thin} {Films}},
	volume = {1},
	language = {en},
	journal = {ACS Appl. Eng. Mater.},
	author = {Zakay, Noy and Lombardo, Luca and Maman, Nitzan and Parvis, Marco and Vradman, Leonid and Golan, Yuval},
	year = {2023},
	pages = {1367=1374},
}

\clearpage

\setcounter{table}{0}
\setcounter{section}{0}
\setcounter{figure}{0}
\setcounter{equation}{0}
\setcounter{section}{0}

\renewcommand\thesection{S\arabic{section}}
\renewcommand\thetable{S\arabic{table}}
\renewcommand\thefigure{S\arabic{figure}}
\renewcommand\theequation{S\arabic{equation}}

\section{Supplemental Information}

\begin{table}[H]
        \caption{\label{tab:table1}DC power supplies corresponding to each target used for combinatorial deposition.}
        \begin{ruledtabular}
        \begin{tabular}{ccc}
        Target&Power Supply\\
        \hline
        Cr&DCXS-750-4 Multiple Sputter Source DC\\
        Fe&Advanced Energy MDX-1k\\
        Nb&DCXS-750-4 Multiple Sputter Source DC\\
        Mo&DCXS-750-4 Multiple Sputter Source DC\\
        Ta&Advanced Energy MDX-1k\\
        \end{tabular}
        \end{ruledtabular}
    \end{table}

   In examining procedures to fit and simulate individual and combinatorial sputter deposition areal mass density profiles, a rotated two-dimensional Gaussian shape was considered:

    \begin{equation*}
         m_i(x, y) = \frac{A}{\sigma_{x}\sigma_{y}2\pi}e^{-\frac{1}{2}\left[a(x-x_{0})^2+2b(x-x_{0})(y-y_{0})+c(y-y_{0})^2\right]},
        \end{equation*}

    \noindent where $a$, $b$, and $c$ account for rotations by an angle $\theta$ as described by:

    \begin{equation*}
         a = \frac{\cos^2(\theta)}{2\sigma^2_x}+\frac{\sin^2(\theta)}{2\sigma^2_y},
        \end{equation*}

    \begin{equation*}
         b = -\frac{\sin(\theta)\cos(\theta)}{2\sigma^2_x}+\frac{\sin(\theta)\cos(\theta)}{2\sigma^2_y},
        \end{equation*}

    \noindent and,

    \begin{equation*}
         c = \frac{\sin^2(\theta)}{2\sigma^2_x}+\frac{\cos^2(\theta)}{2\sigma^2_y}.
        \end{equation*}

    A comparison between fitting parameters and errors for both the rigid and rotated shapes optimized using Mo depositions at four different powers is provided in Table S2. While both shapes provide nearly identical $R$$^2$ values, the additional angular degree of freedom in the rotational Gaussian expression, which significantly varies deposition-to-deposition, produces larger variances in the $\sigma_x$ and $\sigma_y$ parameters. As such, using parameters optimized with a rotatable Gaussian shape are expected to be less accurate at predicting aerial density profiles for arbitrary target powers. Owing to this disadvantage, rigid two-dimensional Gaussian peak shapes were used for all aerial density profile fitting.

    \begin{table}[H]
        \caption{\label{tab:table2}Comparison between mean fitting parameters optimized using rigid and rotatable two-dimensional Gaussian peak shapes for four individual Mo depositions.}
        \begin{ruledtabular}
        \begin{tabular}{ccccc}
        Gaussian Peak Shape&$\sigma_x$ (cm)&$\sigma_y$ (cm)&$\theta$($^\circ$)&$R$$^2$\\
        \hline
        Rigid&5.94 $\pm$ 0.03&5.27 $\pm$ 0.12&--&0.998 $\pm$ 0.001\\
        Rotated&5.50 $\pm$ 0.57&5.51 $\pm$ 0.50&26.5 $\pm$ 15.0&0.997 $\pm$ 0.003\\
        \end{tabular}
        \end{ruledtabular}
    \end{table}

    \begin{table}[H]
        \caption{\label{tab:table3}Optimized parameters for two-dimensional Gaussian shapes fit to four individual depositions of Cr with different applied target powers.}
        \begin{ruledtabular}
        \begin{tabular}{ccccc}
        Elem.&$\sigma_x$ (cm)&$\sigma_y$ (cm)&$x_0$ (cm)&$y_0$ (cm)\\
        \hline
        Cr&5.59 $\pm$ 0.06&5.71 $\pm$ 0.09&4.11 $\pm$ 0.56&-3.51 $\pm$ 0.28\\
        \end{tabular}
        \end{ruledtabular}
    \end{table}
    
    \begin{table}[H]
        \caption{\label{tab:table4}Optimized two-dimensional Gaussian fit parameters for each element from combinatorial sputter depositions of CrFeMoNbTa films at three different uniform powers.}
        \begin{ruledtabular}
        \begin{tabular}{ccccccc}
        Elem.&$\sigma_x$ (cm)&$\sigma_y$ (cm)&$x_0$ (cm)&$y_0$ (cm)&$\mathcal{A}$ (g cm$^{2}$ J$^{-1}$)&$b$ (g s$^{-1}$)\\
        \hline
        Cr&5.57 $\pm$ 0.07&5.67 $\pm$ 0.04&3.34 $\pm$ 0.48&-4.76 $\pm$ 0.19&3.23 $\times$ 10$^{-6}$&-0.28 $\times$ 10$^{-6}$\\
        Fe&5.25 $\pm$ 0.16&5.55 $\pm$ 0.22&0.98 $\pm$ 0.44&4.90 $\pm$ 0.24&2.30 $\times$ 10$^{-6}$&0.40 $\times$ 10$^{-6}$\\
        Mo&5.60 $\pm$ 0.02&4.97 $\pm$ 0.05&5.28 $\pm$ 0.27&1.22 $\pm$ 0.22&5.23 $\times$ 10$^{-6}$&0.09 $\times$ 10$^{-6}$\\
        Nb&5.66 $\pm$ 0.05&5.77 $\pm$ 0.12&-2.79 $\pm$ 0.27&-3.78 $\pm$ 0.32&3.39 $\times$ 10$^{-6}$&-0.29 $\times$ 10$^{-6}$\\
        Ta&5.19 $\pm$ 0.34&5.18 $\pm$ 0.52&-3.72 $\pm$ 0.12&1.34 $\pm$ 0.34&6.61 $\times$ 10$^{-6}$&1.94 $\times$ 10$^{-6}$\\

        \end{tabular}
        \end{ruledtabular}
    \end{table}

    Equation~\ref{eq:generalized} is the generalized form of Equation 7 in the main text, where non-covariance is not assumed for each individual process. In such a case where non-covariance breaks down, the inter-related effects of each process on one another must be appropriately captured by the calibration process for accurate calculation of optimized powers.

    \begin{equation} \label{eq:generalized}
        P_i = \sum_{j}(\boldsymbol{\mathcal{A}}^{-1}_{ij})\left(\frac{\rho_jnC_j}{N_\mathrm{A}\alpha_j(x,y)M}-b_j\right)
                \end{equation} 
\newline

    \begin{figure}[H]
    
        \includegraphics[width=1\textwidth]{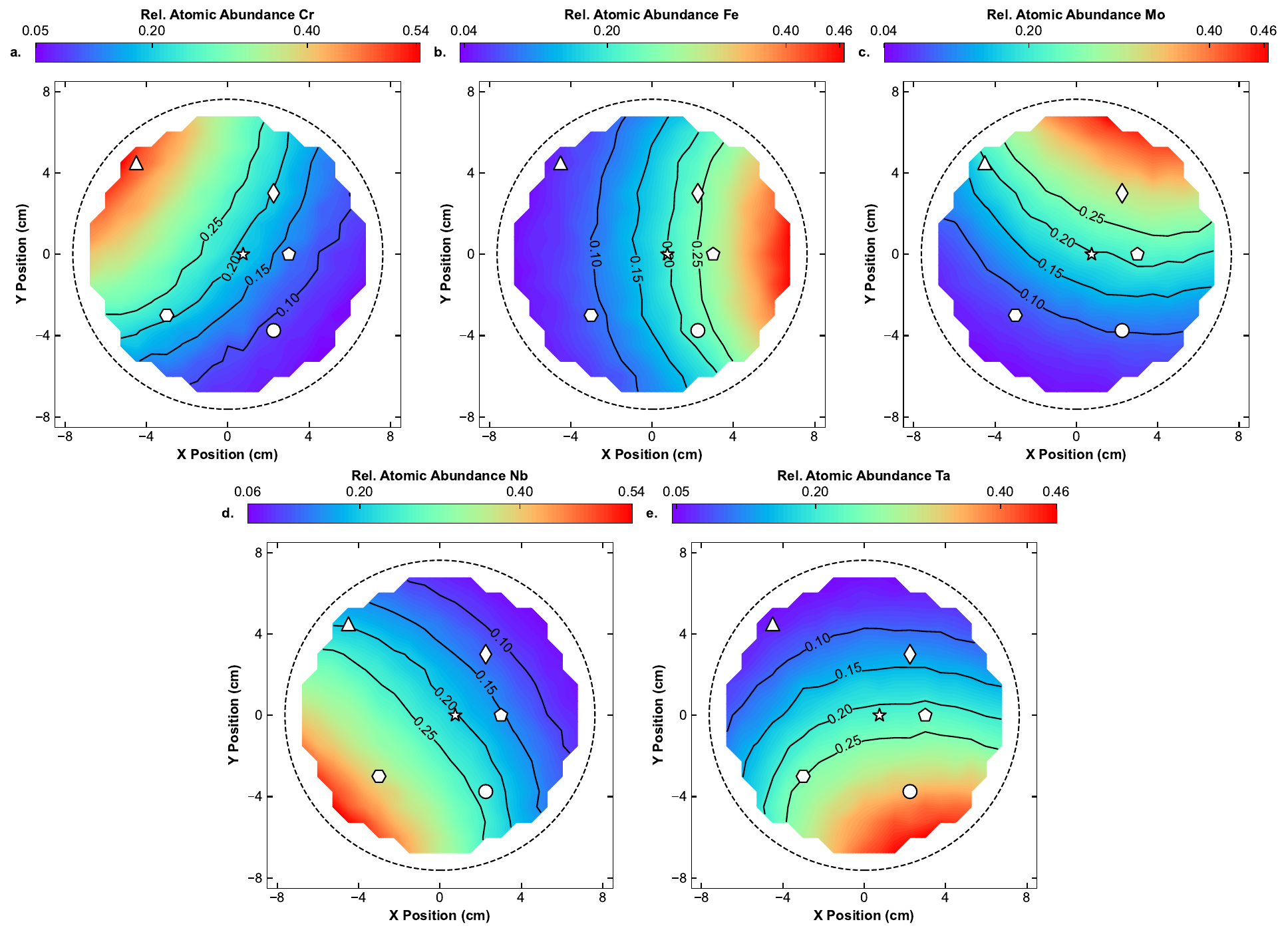}
        \caption{Relative atomic abundances for (a) Cr, (b) Fe, (c) Mo, (d) Nb, and (e) Ta calculated from WDXRF measurements of a combinatorial film deposited using target powers optimized to obtain an equiatomic composition near wafer center. The white symbols on each panel correspond to the positions where EDS composition measurements were made, as detailed in supplemental Fig.~S1. In all panels, warm and cool colors indicate larger and smaller values, respectively, as detailed by each corresponding color bar. The black dotted circle in all panels represents the extent of the 6-inch dimater Si wafer.}

        \label{fig:combi_dep_sigma}

    \end{figure}

    \begin{figure}[H]
    
        \begin{center}
        \includegraphics[width=1\textwidth]{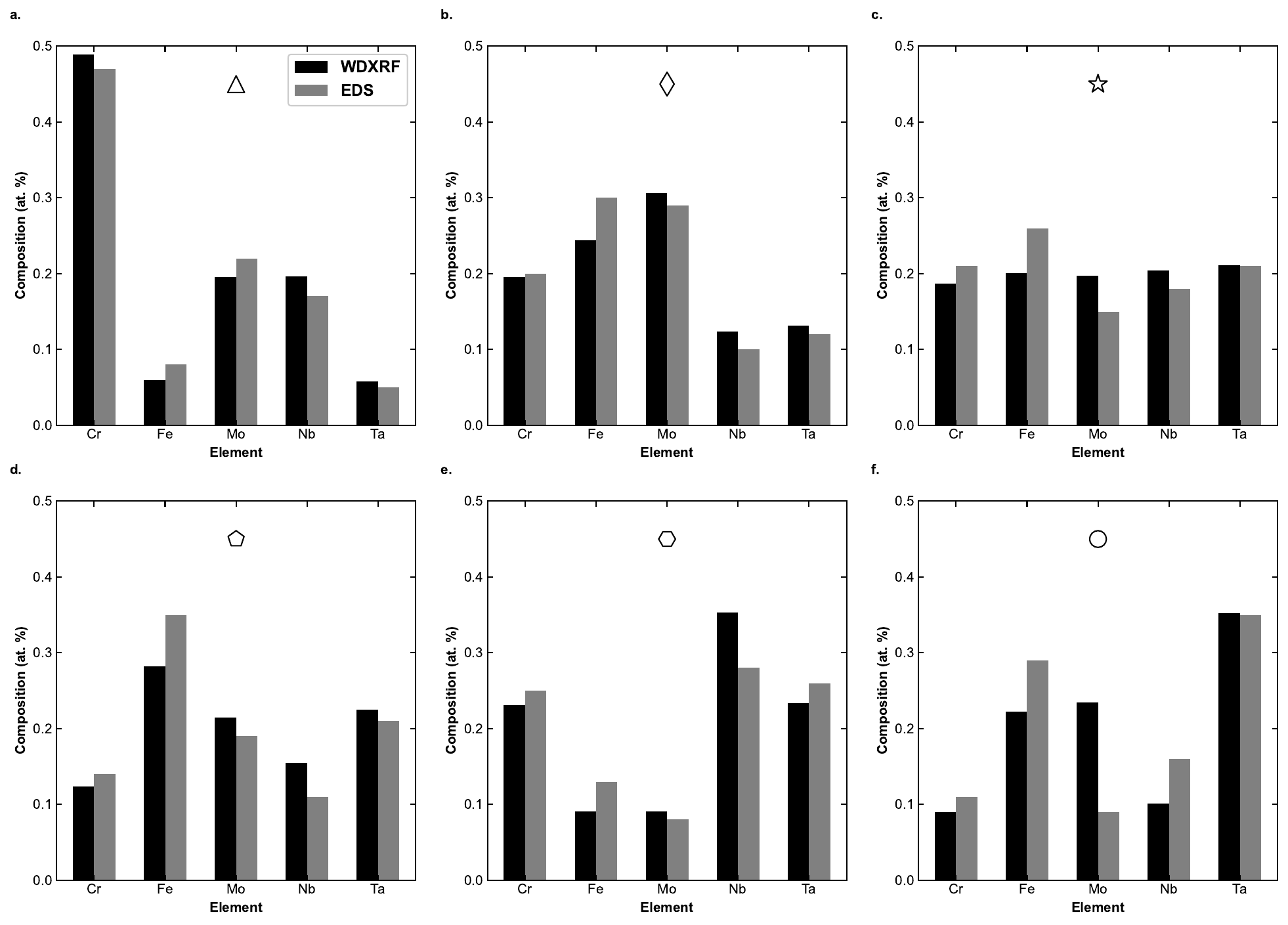}
        \caption{Bar charts comparing Cr, Fe, Mo, Nb, and Ta atomic concentrations measured using WDXRF (black) and EDS (gray). Each panel corresponds to a point on the wafer in Fig.~S1, both annotated by matching symbols.}
        \end{center}
        \label{fig:Scaled_sim}
    \end{figure}

   \begin{figure}[H]
    
        \begin{center}
        \includegraphics[width=1\textwidth]{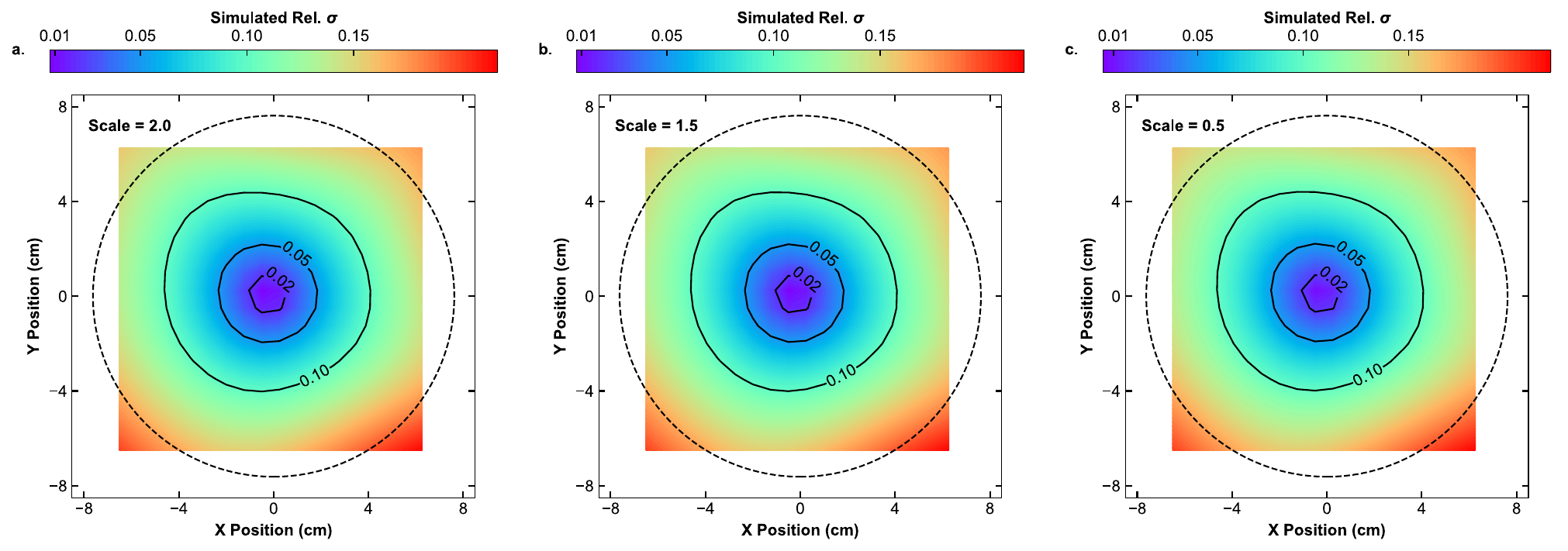}
        \caption{Simulated relative standard deviations of CrFeMoNbTa compositions where powers of 35, 40, 32, 50, and 36 W (from the optimization procedure in the main text) utilized for Cr, Fe, Mo, Nb, and Ta targets, respectively, have been divided by (a) 2, (b) 1.5, and (c) 0.5. Warm and cool colors indicate larger and smaller values, respectively, as detailed by each corresponding color bar, and the power scale factor is annotated in the upper left of each panel. The black dotted circle in all three panels represents the extent of the 6-inch dimater Si wafer.}
        \end{center}
        \label{fig:Scaled_sim}
    \end{figure}

    \begin{figure}[H]
    
        \includegraphics[width=1\textwidth]{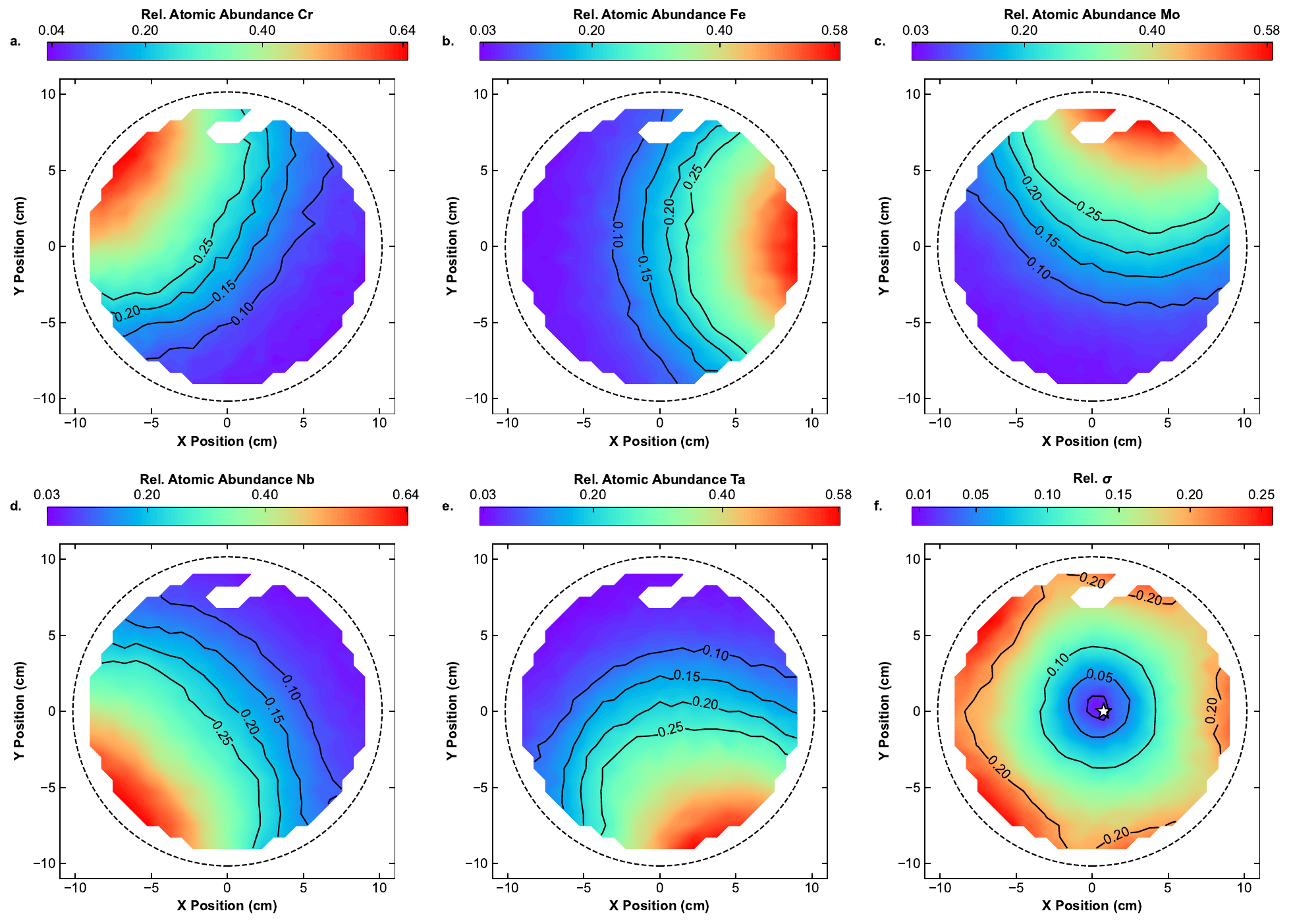}
        \caption{Relative atomic abundances for (a) Cr, (b) Fe, (c) Mo, (d) Nb, and (e) Ta calculated from WDXRF measurements of a combinatorial film deposited on an 8 inch Si substrate using target powers optimized to obtain an equiatomic composition near wafer center. (f) RMS deviation of each elemental composition. The white star near the center of this plot corresponds to the position with the lowest relative $\sigma$ value. In all panels, warm and cool colors indicate larger and smaller values, respectively, as detailed by each corresponding color bar. In these plots, several WDXRF measurement points near the top of the wafer are missing due to visible delamination of the film. The black dotted circle in all panels represents the extent of the 8-inch dimater Si wafer.}

        \label{fig:combi_dep_sigma}
    \end{figure}

\end{document}